# Robust Digital Watermarking Method Based on Adaptive Feature Area Extraction and Local Histogram Shifting

Zi-yu Jiang[1], Chi-Man Pun[1*], Xiao-Chen Yuan[1] and Tong Liu[1]

[1]Department of Computer and Information Science, University of Macau, Macau SAR, China.

*Corresponding author(s). E-mail(s): cmpun@umac.mo;

**Abstract**

A new local watermarking method based on histogram shifting has been proposed in this paper to deal with various signal processing attacks (e.g. median filtering, JPEG compression and Gaussian noise addition) and geometric attacks (e.g. rotation, scaling and cropping). A feature detector is used to select local areas for embedding. Then stationary wavelet transform (SWT) is applied on each local area for denoising by setting the corresponding diagonal coefficients to zero. With the implementation of histogram shifting, the watermark is embedded into denoised local areas. Meanwhile, a secret key is used in the embedding process which ensures the security that the watermark cannot be easily hacked. After the embedding process, the SWT diagonal coefficients are used to reconstruct the watermarked image. With the proposed watermarking method, we can achieve higher image quality and less bit error rate (BER) in the decoding process even after some attacks. Compared with global watermarking methods, the proposed watermarking scheme based on local histogram shifting has the advantages of higher security and larger capacity. The experimental results show the better image quality as well as lower BER compared with the state-of-art watermarking methods.

# 1 Introduction

Nowadays, the copyright protection becomes an important issue because of the rapidly developed communication networks[1–3]. There are many aspects which need digital watermarking to identify copyright such as audio, video and picture. The technology of digital watermarking is practical and useful, becoming a promising method to tackle information copyright [4–8]. The watermarking methods are also various in order to adapt many situations. For example, phase coding was applied on audio watermarking as illustrated in [9] in order to reach the imperceptibility and security. The discrete cosine transform (DCT) was used in [10] to transform the original audio into another domain for watermark embedding process. An echo kernel was analyzed in [11] for audio watermarking in order to reach a lower distortion of the original audio and realize the imperceptibility. In [12], in order to reach the decent image quality and robustness, the DCT was used for watermark embedding. And the coefficients were divided into different groups for watermark decoding. The corresponding watermarking method can also be altered due to the practical matters. The watermarking method used in [13] aimed at signing patient information in brain tumor image. Thus a location map of the image was applied to select non-important area as watermarking area. The important area of the image is the tumor part which would not be affected. In [14], the backward-propagation neural network (BPNN) technique and just-noticeable difference (JND) model were incorporated into a block-wise DCT-based scheme for achieving effective blind image watermarking. This method showed imperceptibility as well as robustness against various attacks which took the consideration of DCT coefficients and the coefficients located in adjacent blocks in order to develop the watermarking scheme. Discrete wavelet transform (DWT) was used in [15, 16] for watermark embedding and extraction from gray-scale image. By applying human visual system (HVS) characteristics in DCT domain during the watermark embedding process, the watermarking algorithm in [15] exhibited good imperceptibility and robustness against many common image processing attacks.

Various image-watermarking methods have been developed over the last decades. Some of them operate the watermark embedding process in the whole image; others focus on local areas of the image for embedding process. Watermark embedding methods are also differing in hundreds of ways. A good image watermarking method should have the ability of imperceptibility, security and robustness. Imperceptibility means that the watermark should be perceptually invisible. Security means that the watermark should not be hacked by hackers. Robustness indicates the correctness of watermark extraction after undergoing different kinds of attacks. There are two types of attacks which are geometric attacks including cropping, scaling and rotation, and signal processing attacks such as compression, noise addition and filtering. Different embedding method has different properties in the aspects of imperceptibility, security and robustness. For example, the method in [17, 18] and [19] were resistant to JPEG compression and median filtering but fragile to rotation. In [20], a circular symmetric watermark embedding method was proposed which processed in 2D Discrete Fourier Transform (DFT) domain. By inserting the watermark in the middle frequency range of the image, this method solved the problem of visible changes in low frequency as well as the compression affects in high frequency under complicated calculating. Circular domain embedding method had also been proposed in [21]. By using Zernike moments and pseudo-Zernike moments, it was robust against image rotation, scaling and flipping but required relatively long-playing time. Thus, fast algorithms should be adopted to further improve calculating speed. Local Zernike moments was used in [22] which were computed over circular patches around feature points of the image. Since corner points could be detected even after image distortions, the Harris corner detector was used to extract feature point of the image in this paper. Some watermarking methods also use SWT to implement the embedding process, like [23] and [24]. They chose coefficients of SWT domain for watermark insertion and the watermark could be a branch of pixels of small image or in binary bits format. The defect of such method is that the watermark format is restricted although it can achieve relatively high imperceptive quality. The work in [25] embedded the watermark in Fourier-Mellin domain by applying Fourier-Mellin transform. This method was resist to scaling and rotation. However, it was vulnerable to cropping. In [26], the Singular Value Decomposition (SVD) combined with DCT transform were used to deal with cropping attacks. In contrast, this watermarking method showed bad performance on signal processing attacks such as compression and noise addition and some geometric attacks such as rotation and scaling. A modulation method in [27] called transportation natural watermarking (TNW) which was used to select different combinations of sub-bands to reach the low BER against JPEG compression. However, the selection of sub-bands for combination was complex and the PSNR was compensated due to higher BER. Histogram shifting based watermarking method in global scale of the image is being popular in some cases. In [28] and [29], two different histogram shifting methods were proposed which were invariant to rotation and scaling, but in some case the watermarking method showed defect under the attacks of median filtering and cropping. Low frequency domain was used for watermarking in [30] and [31] to achieve robustness against common signal processing. Such watermarking methods provided satisfactory performance for geometric deformations and image processing operations, including low-pass filtering, JPEG compression and cropping. In [32], a local histogram shifting method was proposed and the watermark bits were embedded in each circular area of the image. This method was robust against random bending and median filtering, however it was very sensitive to scaling and rotation. By embedding watermark in DFT domain in [33] this method was resistant to rotation and shearing. However the detection rate was not good under JPEG attacks.

There are many types of feature point extraction methods [34–37]. In [38] and [39], local binary pattern (LBP) method was used to recognize feature patterns of the image. Both of them came up with a new method of computing rotation invariant feature recognition. However, two methods of them were focusing on the pattern tracks of texture images, tracking the specific histogram distribution part of the digital image no matter how about the distribution and how many gray levels it has. Davarzani et al.[40] proposed an improved local binary pattern feature selection method. It could be scale and rotation invariant when recognizing texture descriptions of the digital image. But this method may not be robust under attacks such as compression and cropping. The scale-invariant feature transform (SIFT) feature was used in [41] and [42] which showed good performance to pattern recognition even though the searching patterns have different sizes and orientations. However, the extracted feature points cannot be invariant because when there is any attacks applied on the image, the image pixels would be modified and the SIFT accounting results would be inaccurate in finding the same patterns. In addition, the SIFT is time consuming to compute and recognize a certain number of feature points. Rotation, scaling and translation-invariant (RST-invariant) feature extraction method was proposed in [43] to promote the feature recognition performance with respect to some kinds of attacks such as rotation, scaling and translation. However, since the spatial domain has limited robust features for resisting RST attacks, this method should compute both in spatial domain and in frequency domain, which would be relatively complicated in order to find some specific features. Gradient location-orientation histogram (GLOH) is another type of feature extraction method as it was used in [44]. It used a log-polar location grid to generate several location bins and computed them for feature extraction. And this feature extraction method showed better performance than SIFT especially under illumination changes. However, GLOH could be a more time consuming and complicated method as discussed in [45]. Daisy, an efficient feature recognizing method was proposed in [46] and it showed decent feature detection effect. Daisy is a type of improved gradient computing method which can be computed much faster. Since this feature extraction method is based on pixel gradient computing, it shows rotation-invariant performance and robustness under different attacks. In this paper, we propose a local histogram shifting watermarking method that embeds watermark sequence into several local squares of the digital image. By embedding watermark in local areas of the image, the high imperceptibility can be achieved and a relatively higher PSNR can be obtained through this method. Embedding in local areas makes the hacking more difficult. In addition, by using a secret key to indicate the embedding bands of the histogram in the histogram shifting process, the security to resist unlawful detection and tampering will be improved. An adaptive feature selection method based on Daisy descriptors is proposed to extract the feature points. It selects the local areas which have relatively more gradient changes and more gray levels of the image histogram to ensure the high capacity of each local area and the implementation of embedding procedure. Although the concepts of SWT based noise removal and histogram shifting based watermark embedding have been proposed, the proposed method in this paper is different from the existing schemes. In [23] and [24], the SWT was applied to decompose the input image, and the corresponding low frequency coefficients were used to embed and extract the watermark before reconstruction; while in our method, we apply the SWT to remove the noise and reconstruct the image before embedding and extraction. In [30], the Gaussian filtering was applied for noise removal before embedding, however, in this way, the compensation of the high frequency needs be added before watermark embedding to guarantee the success of watermark extraction; while in our method, we don't need to calculate the compensation. For the histogram shifting based watermarking method, in the existing methods, e.g. [30], no more than six gray levels were used to embed one bit, while in our method, 12 gray levels are chosen to embed one bit, which would cause less modification intensity in the histogram and build smoother histogram of the embedded image. The improvement can help to enhance robustness against some attacks such as wave bending, median filtering, and jittering.

In the following section 2, the embedding process of the proposed watermarking method is explained in detail. After the embedding process, we introduce the decoding process in section 3. Section 4 illustrates the experimental results of our method and shows the comparison between our method and state-of-art watermarking methods. After the analyses of the performance corresponding to robustness, capacity and image quality of the proposed method, we make the conclusion and clarify the advantages of the proposed method in section 5.

## 2 Embedding Process of the Proposed Watermarking Method

The flowchart of the proposed watermark embedding process is shown in Fig.1. The whole process can be divided into two sections: adaptive feature areas extraction and histogram shifting based watermark embedding. The process will be descripted in detail in the following two sections. Firstly, adaptive feature points which are related to pixel gradient are extracted by using Daisy descriptor. Then the feature areas are confirmed according to the distribution of the feature points. After the confirmation of local areas, SWT and inverse stationary wavelet transform (iSWT) is implemented on each local area for noise removal in order to get smooth areas for watermark embedding. Afterwards, we implement the histogram shifting based watermark embedding in local areas, local area recovery and watermarked image reconstruction. The whole process aims at selecting local areas which have more histogram gradient for large capacity and embedding watermark into smooth part of local areas by histogram shifting for higher security as well as higher image quality.

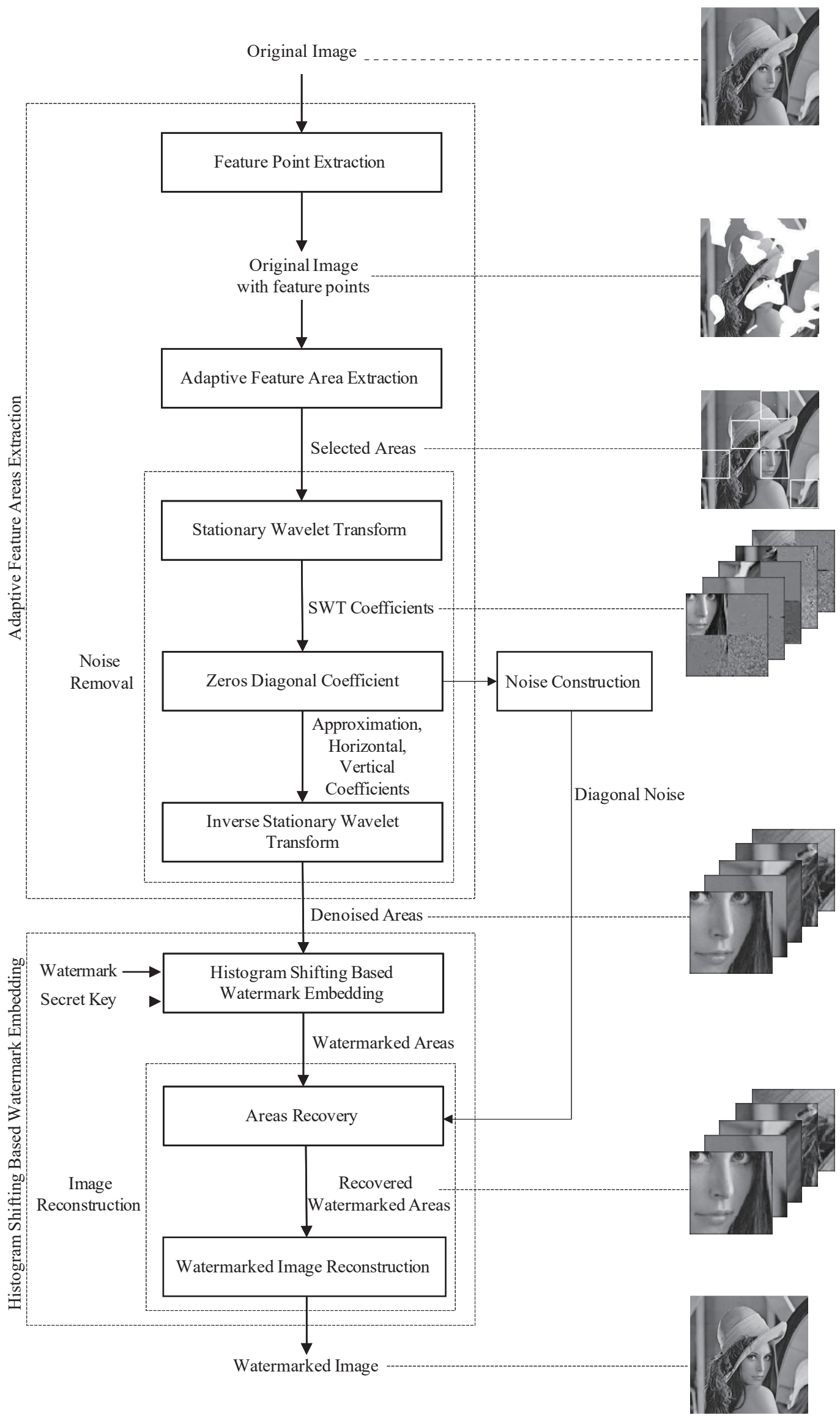

**Fig. 1** Flowchart of watermark embedding process

## 2.1 Adaptive Feature Areas Extraction

We propose an adaptive feature areas extraction method to adaptively select areas for watermark embedding. This method includes three steps: Feature Point Extraction Using Daisy Descriptor, Adaptive Feature Areas Extraction and Noise Removal Using Stationary Wavelet Transform. The first step aims at selecting feature points which fulfil more gradient variance for the large capacity of watermarking. Then, the adaptive feature areas are extracted based on the distribution of feature points. Finally, we remove the noise of the selected areas through SWT implementation and obtain the smooth part of those feature areas for watermark embedding.

### 2.1.1 Feature Point Extraction Using Daisy Descriptor

In the proposed watermarking method, we use Daisy descriptor as the feature extraction method to compute the feature points of the image for the selection of local areas for watermark embedding. Daisy becomes a very fast and efficient way to compute the pixel gradients of the whole image. The structure of the Daisy descriptor is like a daisy flower which has several layers and directions. Unlike SIFT and GLOH descriptors which involve 3D histograms and corresponds to image spatial dimensions as well as the additional dimension to the image gradient direction, Daisy implements the computing process by using several Gaussian filters to realize convolutions of the gradients in specific directions of the pixel which can be done faster [46]. Besides, the Daisy descriptor has a circular grid format with several layers and directions. As described in [46], each circle in Daisy descriptor represents a region where the radius is proportional to the standard deviations of the Gaussian kernels. In feature extraction work, the goal is to find local areas where more pixel gray level variance can be found in the whole image. By using Daisy descriptor to compute image gradient at every pixel location of the image, the histogram would be computed in each region which is centralized at specific pixel location.

Given an image $I$ with pixel number of $M \cdot N$, then $j, j = M \cdot N$ descriptors would be computed. The Daisy descriptor computing principle is described as follows. Firstly, the image gradient norm at location $(m, n)$ for direction $o$ is computed as $G_o(m, n)$. Then, the orientation map which is used to construct the descriptor can be written as

$$G_i = \left(\frac{\partial I}{\partial i}\right)^+ \tag{1}$$

where $i$ is the orientation deviation while $(.)^+$ is the operator, function as $(a)^+ = max(a, 0)$. In this case, each orientation map is then convolved several times with different Gaussian kernels in order to achieve convolutions as well as the reduction of computational complexity. This procedure has been detailed illustrated in [46] . If we compute a Daisy descriptor with $i$ direction and $L$ layers, the descriptor can be formally defined as

$$\begin{aligned} D_j(m_0, n_0) = [ & d_{G^\sigma_1}(m_0, n_0), \\ & d_{G^\sigma_1}(m_0, n_0, R_{11}), \cdots, d_{G^\sigma_1}(m_0, n_0, R_{1i}), \\ & d_{G^\sigma_2}(m_0, n_0, R_{21}), \cdots, d_{G^\sigma_2}(m_0, n_0, R_{2i}), \\ & \cdots \\ & d_{G^\sigma_L}(m_0, n_0, R_{L1}), \cdots, d_{G^\sigma_L}(m_0, n_0, R_{Li})] \end{aligned} \tag{2}$$

where $d_{G^\sigma_L}(m_0, n_0, R_{Li})$ indicates the computed image gradient norm vector at the location with distance $R_L$ from $(m_0, n_0)$ of the direction $i$. The Gaussian kernel is $G^\sigma_l = \frac{R_l(l+1)}{2L}$ where $l$ represents the $lth$ layer in the circular grid of the descriptor and $l=1:L$. We obtain a sequence of numbers exhibiting the normalized histogram of each pixel location relating to the neighborhood of the given pixel, coming from the orientation map [47].

Due to the circular grid format and computation method of Daisy descriptor, it shows the robustness to a variety of attacks and common image processing operations such as noise addition, JPEG compression and rotation, etc. It means the descriptor of specific pixel can still be validly computed even the image is blurred and rotated. Because of such characteristic, we choose Daisy descriptor as the feature pixel selection method. Thus, we can detect corresponding pixels even the image goes through attacks that ensures the stability of local adaptive feature areas extraction.

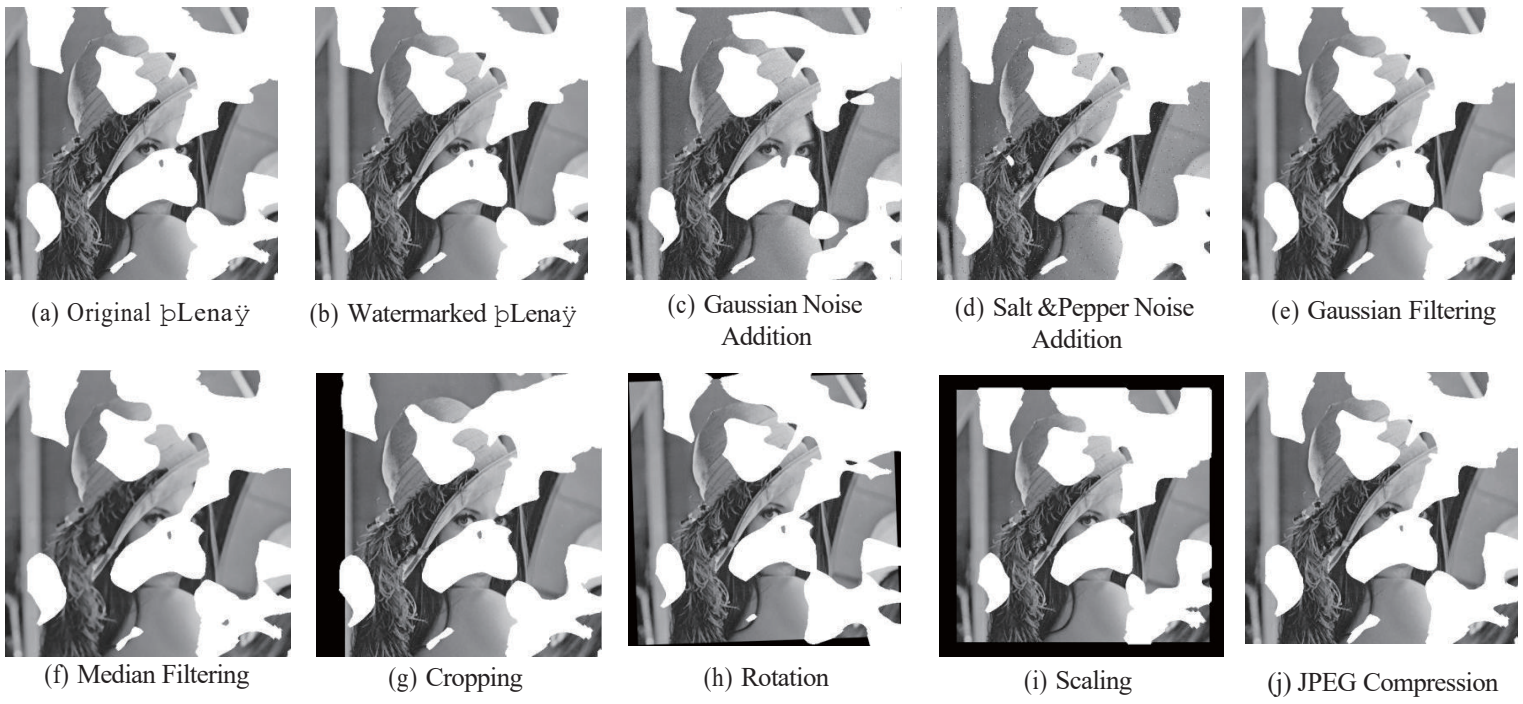

**Fig. 2** Invariance property of feature points selection using Daisy descriptor

After Daisy descriptor has been computed we calculate the summation of each direction of $D_j(m_0, n_0)$ in order to find the point which has the most gradient variance in a specific direction.

$$D_{sumi}(m_0, n_0) = \sum_{l=1}^{L} d_{G^\sigma_L}(m_0, n_0, R_{li}) \tag{3}$$

$$D_{sum}(m_0, n_0) = [D_{sum1}, D_{sum2}, \cdots, D_{sumi}] \quad (4)$$

where $D_{sumi}$ is the gradient summation in direction corresponding to the location $(m_0, n_0)$. Take grayscale image Lena for example, the selected feature points are the white dots which formed areas on the image as shown in Fig. 2. The feature points can be extracted in the watermarked image as well as attacked watermarked image which ensures the watermark decoding procedure can be implemented correctly. The watermark capacity is related to the histogram distribution. As we know that, more gray level of the histogram in the image leads to higher capacity, that is, we can embed more binary bits in the image. In order to obtain more gray levels in the histogram, the local regions which have the most gradient variation would be selected. Thus, more binary bits can be embedded into the digital image and higher image quality can be achieved by implementing our proposed watermark embedding method in such areas.

### 2.1.2 Adaptive Feature Areas Extraction

In this part, adaptive feature areas are extracted according to feature points [48]. Since the selected feature points are those which have more gradients in the neighborhood, we square those feature pixel regions to form feature areas which have more histogram changes. The watermark embedding capacity is related to the image histogram variety, that is, more gray levels in the histogram means larger capacity available. Since we want to reach a larger capacity and embed more binary bits into local areas, those areas with more feature points are extracted. These feature points are selected and grouped into several regions. The grouping method is based on pixel distance and local embedding area radius.

As we can see in Fig. 3, the selected pixels are signed in black and applied on the original image. By computing the pixel areas with morphological processing, we obtain some feature areas. In order to realize the stability of the selection of areas, those areas are adjusted into default image grid according to the nearest principle. Then, selected square areas are obtained. After that, we compute the entropy of each local square area and finally select square areas with the most entropy for watermark embedding.

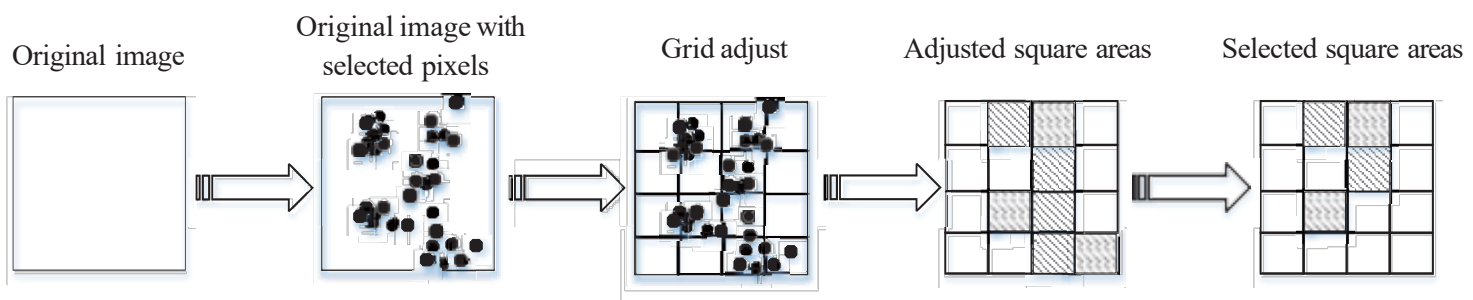

**Fig. 3** Adaptive feature areas extraction

Assume that all selected pixels of the original image $I$ are $P = \{p_1, p_2, \cdots, p_x\}$, then grouped them into $n$ areas

$$x_n = [p_{n1}, p_{n2}, \cdots, p_{nm}] \tag{5}$$

$$X_n = [x_1, x_2, \cdots, x_n] \tag{6}$$

where $n1, n2, \cdots, nm$ indicate the $m$ pixel numbers in $nth$ area and $x_n$ indicates the $nth$ area.

Finally, after comparing the entropy of each grouped area, the selected $s_n$ square areas can be written as

$$A_{s_n} = [p_{s_n 1}, p_{s_n 2}, \cdots, p_{s_n num}] \tag{7}$$

$$A = [A_{s_1}, A_{s_2}, \cdots, A_{s_n}] \tag{8}$$

where $num$ indicates the total number of pixels in the specific selected square area and each selected area has the same number of pixel. The number of pixel depends on the length of a side of the square. And $s_n = 1, 2, 3, \cdots, n$.

### 2.1.3 Noise Removal Using Stationary Wavelet Transform

Common signal attacks can have distinct impacts on image detailed parts, but have less impact on low-frequency parts of the image. Thus embedding watermark in relatively smooth component of the image can achieve the robustness of the watermark. In our method, the denoised areas are used for embedding to ensure the robustness. After the noise removal implementation, the smooth part of the image can be obtained for watermark embedding. The embedded watermark can be correctly detected even after some geometric attacks as well as common image processing operations due to the noise removal implementation. After local area selection, we implement SWT on each local area and zero the diagonal coefficient when doing the iSWT implementation. This procedure allows us to generate the denoised local areas. On the one hand, dislodging diagonal coefficients can wipe off the noise part of the image. On the other hand, by reapplying this diagonal coefficient on the watermarked smooth image, a good image quality can be achieved and the impact of embedding process can be reduced. Compared with Gaussian filter smoothing process, the use of SWT for denoising can be implemented more convenient since we can add noise part directly to the watermarked smooth part without any modification of the noise part. Moreover, at the decoding procedure, the same denoising process can be applied to get the watermarked smooth part. However, the high frequency part processed by Gaussian filter needs to be modified according to the modification of watermarked smooth part; otherwise, the watermarked smooth part cannot be obtained similarly at the decoding end. Thus, SWT and iSWT are used in noise removal step and watermarked areas recovery step in order to ensure the robustness as well as good image quality.

When the selected feature areas are extracted, SWT is applied to decompose the $s_n$ areas into four parts: approximation component, horizontal component, vertical component and diagonal component as shown in equation (9).

$$A_{s_n}^{SWT} = [A_{s_n}^{SWT}(app), A_{s_n}^{SWT}(hor), A_{s_n}^{SWT}(ver), A_{s_n}^{SWT}(dia)] \tag{9}$$

$$A^{SWT} = [A_{s_1}^{SWT}, A_{s_2}^{SWT}, \cdots, A_{s_n}^{SWT}] \tag{10}$$

where $A_{s_n}^{SWT}(app)$ indicates the approximation coefficient after the implementation of SWT of the local area $A_{s_n}$. And the signs *hor*, *ver* and *dia* indicate the SWT coefficients of horizontal, vertical and diagonal respectively. It is known that the diagonal component can be seen as noise. Thus, as shown in equation (11), the three components: approximation, horizontal and vertical components are chosen to do the iSWT to form the denoised area by zeroing diagonal component.

$$A_{s_n}^{SWT} = [A_{s_n}^{SWT}(app), A_{s_n}^{SWT}(hor), A_{s_n}^{SWT}(ver), 0] \tag{11}$$

$$A^{denoised} = iSWT[A_{s_1}^{SWT}, A_{s_2}^{SWT}, \cdots, A_{s_n}^{SWT}] \tag{12}$$

After this procedure, the denoised areas are prepared for embedding. The diagonal coefficient is also being preserved for the use of local embedded areas recovery in later process.

## 2.2 Histogram Shifting Based Watermark Embedding

We use histogram shifting method to implement watermark embedding because this method can be implemented directly in the spatial domain which has the advantages of convenience and imperceptibility. Compared to other embedding domain such as the frequency domain, histogram shifting method can be more efficient and time-saving. The watermark capacity relies on the number of effective gray level of the histogram. Embedding the watermark in the whole image that is in global span, the histogram of the image can be used only once to insert a specific number of binary bits as watermark. In order to conquer the drawback of restricted capacity of global embedding method, we propose the local embedding method which allows many bunches of binary bits to embed in many different local areas of the image. Thus the histogram can be used effectively since each local area can generate a unique local histogram. In addition, by embedding watermarks into different local areas, the security is enhanced correspondingly since the hacker cannot determine the specific embedding location.

### 2.2.1 Gray Level Grouping

In order to embed watermark into the image, we use histogram shifting method to realize this process. We can embed watermark bit ‘0’ or ‘1’ through transferring a specific number of pixels from one histogram part to another in order to generate comparable pixel numbers between two histogram parts, and extract watermark bit by comparing the pixel numbers of two histogram parts. We implement gray level grouping in each histogram of the local area, forming gray level bins and groups which are used for watermark embedding. The gray level grouping procedure is shown as Fig. 4.

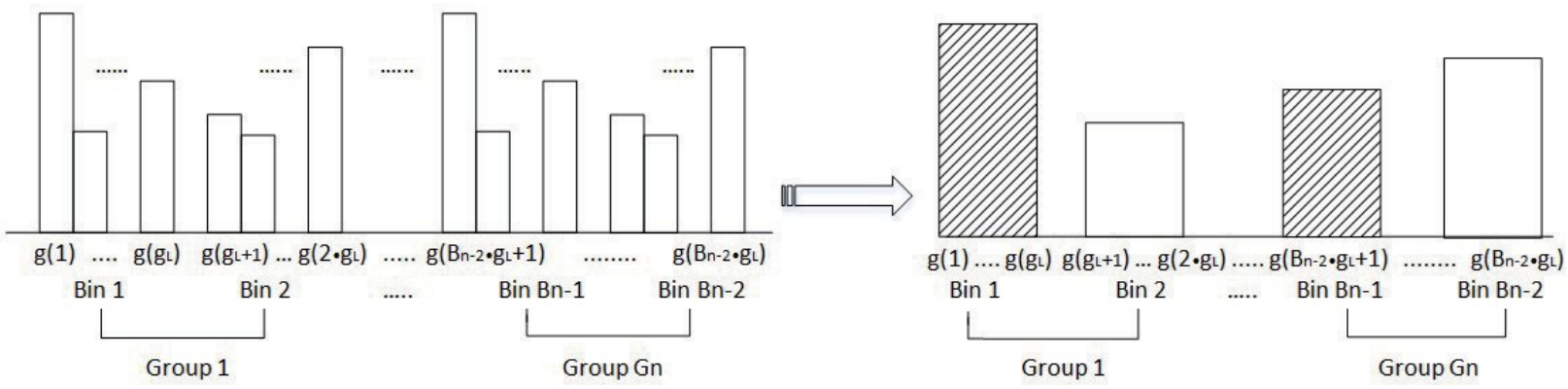

**Fig. 4** Gray level grouping procedure

Firstly, group all gray levels into $B_n$ bins by successively grouping the number of $g_l$ levels together and form them into one bin. Assume that the de-noised local square area $I'$ has $G$ gray levels, for example, an 8-bit gray scale image has $G = 256$ gray levels, ranging from *g(1)* to *g(256)*. The histogram of an image indicates the number of pixels of each gray level. In the proposed method, we divide $G$ gray levels into $B_n$ bins:

$$B_n = \lfloor \frac{G}{g_l} \rfloor \tag{13}$$

where $g_l$ means the number of gray levels in each bin, $\lfloor . \rfloor$ is the floor function. In this case, the number of pixels in *mth* bin *Bin(m)* can be illustrated as

$$N_{bin}(m) = g((m-1) \cdot g_l + 1) + g((m-1) \cdot g_l + 2) + \cdot \cdot \cdot + g(m \cdot g_l) \tag{14}$$

where $m = 1, 2, \cdot \cdot \cdot, B_n$ and *g(x)* indicates the pixel number in gray level $x$ in which $x = ((m-1) \cdot g_l + 1) : (m \cdot g_l)$. Secondly, group all bins into groups by successively grouping two bins into one group. The number of formed groups can be shown as

$$G_n = [\frac{B_n}{2}] \tag{15}$$

Where $G_n$ indicates the total number of formed groups and [.] is the round function. The number of pixels in *dth* group *Group(d)* can be illustrated as

$$N_{group}(d) = g((d-1) \cdot g_l + 1) + g((d-1) \cdot g_l + 2) + \cdot \cdot \cdot + g((d+2) \cdot g_l \tag{16}$$

Where $d = 1, 2, \cdot \cdot \cdot, G_n$ and *g(x)* indicates the pixel number in gray level $x$ in which $x = ((d-1) \cdot g_l + 1) : ((d+2) \cdot g_l)$. Obviously, more pixels in group, higher robustness can be achieved since more pixels can be shifted between the bins of the group. In addition, more groups mean larger capacity due to the embedding principle that one group can provide a capacity of one binary bit. The secret key which is a binary sequence in the algorithm is used to indicate whether the corresponding group is watermarked or not as shown in

(17)

$$K = [K(1), K(2), K(3), \cdots, K(G_n)] \tag{17}$$

The length of the secret key equals the number of formed groups of one local area. Since more than one local area would be chosen for embedding, the key for each local area would be generated by permutation function at the embedding stage. If we use '1' to indicate the embedding group, the summation of the key equals the length of the embedded watermark bit sequence $len_w$. The groups for watermark embedding can be illustrated as:

$$\begin{aligned} Group_{embed}(emb) =& [Group(1), Group(2), \cdots, Group(G_n)] \\ & \cdot [K(1), K(2), K(3), \cdots, K(G_n)] \end{aligned} \tag{18}$$

where $emb = 1, 2, \cdots, len_w$.

### 2.2.2 Watermark Embedding by Histogram Shifting

Histogram shifting method based watermark embedding has the advantages of visual imperceptibility and good image quality. Since histogram shifting is implemented by transforming pixel from one gray level to another, it can be implemented directly in the spatial domain which is an efficient and convenient method for watermark embedding. In our method, we group gray levels into bins and transform a specific number of pixels from one bin to another. The distance of pixel modification is controlled in a small range which ensures the visual imperceptibility as well as good image quality.

In our embedding method, we embed watermark bit sequence $w$ into selected groups and insert one bit in each group.

Assume that $w_1, w_2, \cdots, w_{len_w}$ is the watermark bit sequence which would be embedded into $len_w$ chosen group. It is known that each group has two bins that allow us to transfer pixels through one bin to another. The watermark embedding rule, take $dth$ group for example, is

$$\begin{cases} \frac{N_{bin}(2d-1)}{N_{bin}(2d)} \geq S, & w_d = 1 \\ \frac{N_{bin}(2d-1)}{2d} \leq \frac{1}{S}, & w_d = 0 \end{cases} \tag{19}$$

where $S$ is the watermark strength which influences the number of transformed pixel. And $d$ indicates the group number. The above mentioned embedding rule indicates that if $w_d = 1$, a certain number of pixels $N_1$ from Bin 2 would be transferred to Bin 1 in the specific group; if $w_d = 0$ , a certain number of pixels $N_0$ from Bin 1 would be transferred to Bin 2 as shown in Fig.5. Thus we can compute the transferring number of pixels in each condition. The minimum $N_1$ and $N_0$ are indicated as:

$$\begin{cases} N_1 = \frac{S \cdot N_{bin}(2d) - N_{bin}(2d-1)}{S+1} \\ N_0 = \frac{S \cdot N_{bin}(2d-1) - N_{bin}(2d)}{S+1} \end{cases} \tag{20}$$

where $N_1$ and $N_0$ indicate the transformed number of pixels from one bin to another in the same group. And $d$ is the group number while $S$ is the watermark strength. $N_{bin}(2d - 1)$ indicates the pixel number in the Bin 1 of group $d$ while $N_{bin}(2d)$ indicates the pixel number in the Bin 2 of group $d$.

After such embedding procedure, the watermarked de-noised square images $I'_{wi=1:n}$ are obtained. The local histogram shifting method is shown in Fig.5. In this case, six gray levels are chosen to form one bin and two bins are chosen to form an embedding group which has a capacity of one bit. If the watermark bit equals 0, the pixels in Bin 1 would be transferred into Bin 2 while if the watermark bit equals 1, the pixels in Bin 2 would be transferred into Bin 1.

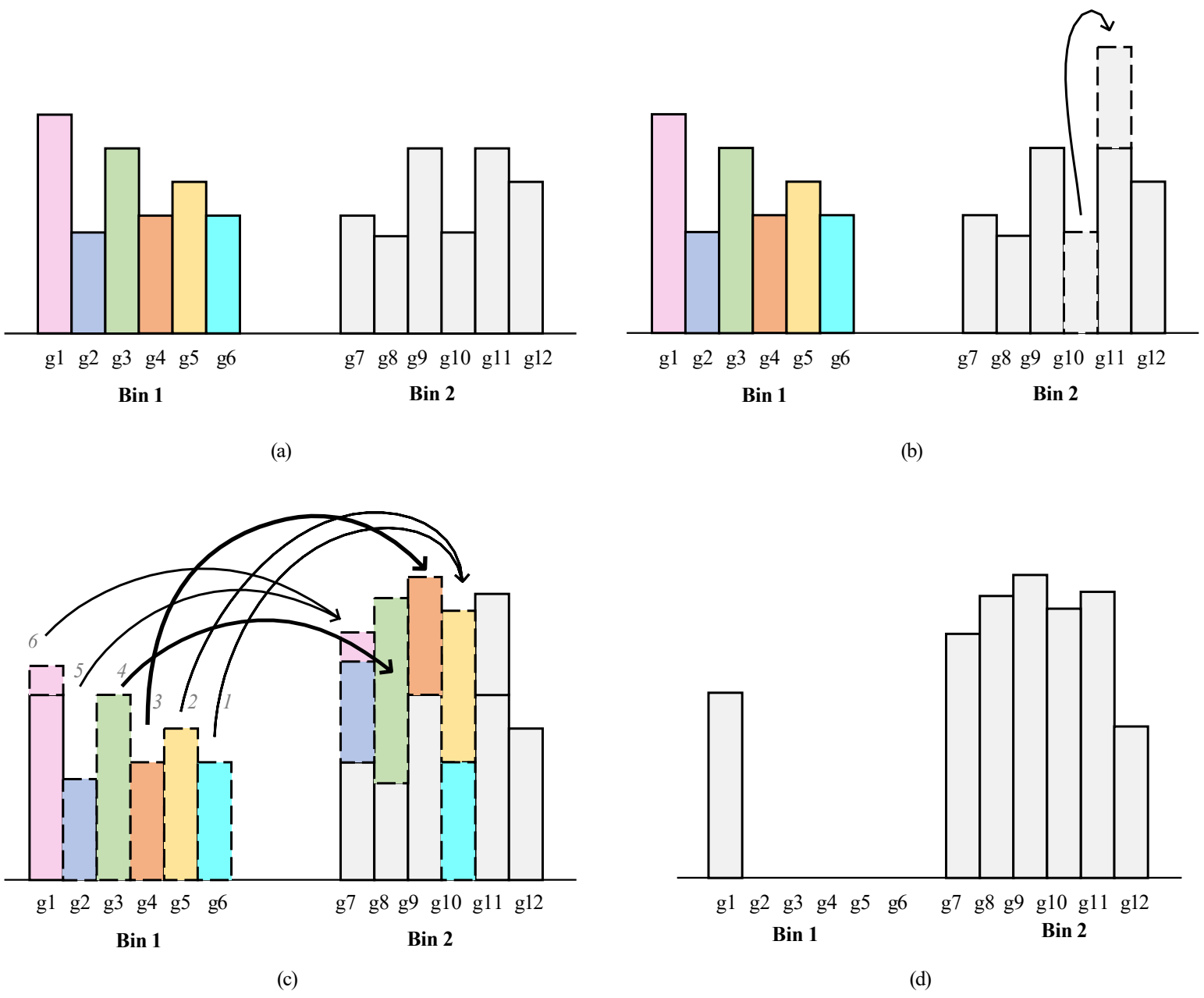

**Fig. 5** Illustration of transferring pixels between Bin 1 and Bin 2

In order to avoid too many pixels in the same gray level, we firstly transfer a number of pixels from g10 to g11 as shown in Fig. 5 (b) since g10 will receive the pixels transferred from g5 and g6. And the number of pixels in g11 will not be concerned about the exceeded number because after pixels transfer between two bins, the gray level distribution of transformed bin will be comparative and the peak-like things will not appear which could be scappled during the iSWT denoising process.

We use the sign $N$ (g) to indicate the pixel number in g*th* gray level. Take embedding watermark bit '0' for example, the pixel transfer approach can be described as follows:

Case 1: If $N$ (g10)≥ $N_0$, more $N_0$ pixels from gray level g10 to gray level g11 within Bin 2. Then, select $N_0$ pixels from Bin 1 as follows. The shifting process is indicated as Fig. 5 (c).

1) if $N$ (g6)≥ $N_0$, choose all $N_0$ pixels from gray level g6 in Bin 1, move them to gray level g10 in Bin 2. Then the histogram shifting process is completed.
2) if $N$ (g6)< $N_0$, choose $N$ (g6) pixels from gray level g6 in Bin 1, move them to gray level g10 in Bin 2. Then, if $N$ (g5)≥ ($N_0$-$N$ (g6)), choose $N_0$-$N$ (g6) pixels from gray level g5 in Bin 1, move them to gray level g10 in Bin 2. The histogram shifting process is completed.
3) if $N$ (g5)< ($N_0$-$N$ (g6)), choose $N$ (g5) pixels from gray level g5 in Bin 1, move them to gray level g10 in Bin 2. Then, if $N$ (g4)≥ ($N_0$-$N$ (g6)-$N$ (g5)), choose $N_0$-$N$ (g6)–$N$ (g5) pixels from gray level g4 in Bin 1, move them to gray level g9 in Bin 2. And the histogram shifting process is completed.
4) if $N$ (g4)< ($N_0$-$N$ (g6)-$N$ (g5)), choose $N$ (g4) pixels from gray level g4 in Bin 1, move them to gray level g9 in Bin 2. Then, if $N$ (g3)≥ ($N_0$-$N$ (g6)-$N$ (g5)-$N$ (g4)), choose $N_0$-$N$ (g6)-$N$ (g5)-$N$ (g4) pixels from gray level g3 in Bin 1, move them to gray level g8 in Bin 2. And the histogram shifting process is completed.
5) if $N$ (g3)< ($N_0$-$N$ (g6)-$N$ (g5)-$N$ (g4)), choose $N$ (g3) pixels from gray level g3 in Bin 1, move them to gray level g8 in Bin 2. Then, if $N$ (g2)≥ ($N_0$-$N$ (g6)-$N$ (g5)-$N$ (g4)-$N$ (g3)), choose $N_0$-$N$ (g6)-$N$ (g5)-$N$ (g4)-$N$ (g3) pixels from gray level g2 in Bin 1, move them to gray level g7 in Bin 2. And the histogram shifting process is completed.
6) if $N$ (g2)< ($N_0$-$N$ (g6)-$N$ (g5)-$N$ (g4)-$N$ (g3)), choose $N$ (g2) pixels from gray level g3 in Bin 1, move them to gray level g7 in Bin 2. Then, if $N$ (g1)≥ ($N_0$-$N$(g6)-$N$(g5)-$N$(g4)-$N$(g3)-$N$(g2)), choose $N_0$-$N$(g6)-$N$(g5)-$N$(g4)-$N$(g3)-$N$(g2) pixels from gray level g1 in Bin 1, move them to gray level g7 in Bin 2. And the histogram shifting process is completed.
7) if $N$ (g1)< ($N_0$-$N$ (g6)-$N$ (g5)-$N$ (g4)-$N$ (g3)-$N$ (g2)) , choose $N$ (g1) pixels from gray level g1 in Bin 1, move them to gray level g7 in Bin 2. Then the histogram shifting process is completed.

Case 2: If $N$ (g10)< $N_0$, move all $N$ (g10) pixels from gray level g10 to gray level g11 within Bin 2. Then, the procedure of selecting $N_0$ pixels from Bin 1 and transforming them into Bin 2 is the same as Case 1.

In this approach, most selected pixels are transformed into the near gray level within the same group. This means the modification of the pixel value is small and the good perceptual quality as well as the PSNR of the watermarked area can be realized. In addition, dispersedly transforming pixels from one bin to another can form the relatively smooth histogram which cannot be easily scappled by denoising function as well as image attacks.

### 2.2.3 Watermarked Image Reconstruction

In order to enhance the image quality of watermarked image, the reconstruction process is implemented by re-put the diagonal coefficient of SWT function on the watermarked local areas. After that, we replace the local areas on the original image to generate the watermarked image with local embedding modification.

After the watermarking process, we obtain the denoised watermarked areas $A^w$.

$$A^w = \{A^w_1, A^w_2, \cdots, A^w_n\} \tag{21}$$

where $A^w_n$ is the $nth$ denoised watermarked area. In order to increase the image quality and avoid PSNR degradation, the noise part of diagonal coefficient of SWT should be re-applied on each watermarked local area. This procedure can be easily implemented by making the subtraction between the original areas and the denoised areas to get the noise parts of the areas $A^{Nois}_n$. Then make the addition of these noise parts and the watermarked areas, getting the constructed areas $A_{construct}$.

$$A_{construct} = \{A^w_1 + A^{Nois}_1, A^w_2 + A^{Nois}_2, \cdots, A^w_n + A^{Nois}_n\} \tag{22}$$

Finally, each watermarked square image is re-applied on the original position of the whole image $I$ and the watermarked image $I_w$ is obtained. This process is shown as Fig. 6.

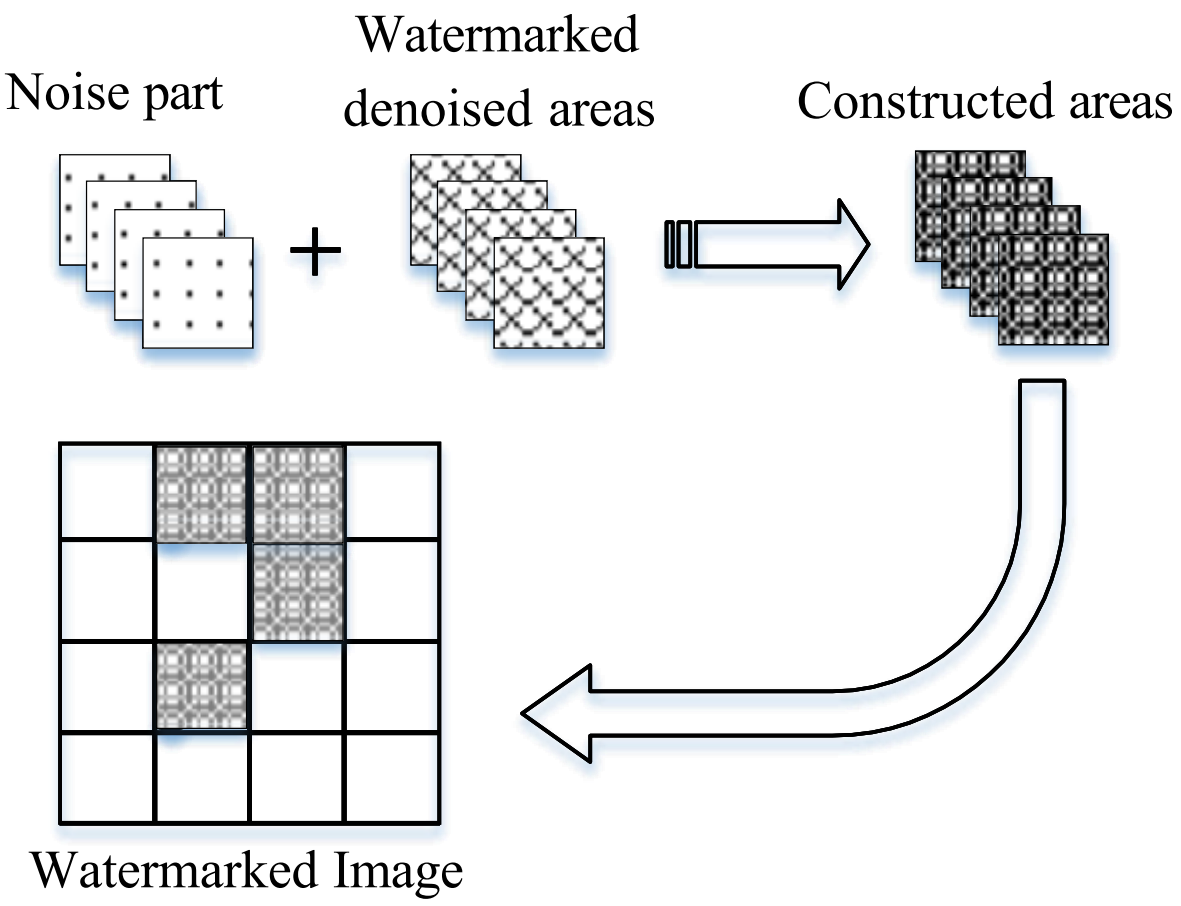

**Fig. 6** Watermarked Image Reconstruction process

# 3 Watermark decoding process

Before the watermark extraction procedure, a rotation angle estimation process is introduced in case of rotation attack. Since in the proposed method, the local areas for embedding are squares, the regular position of the decoded image should be achieved by angle adjusting. In angle estimation process, the binarization process is firstly applied on the received image; afterwards, the altered angle of the obtained image is calculated by using the deviation distance of the image edge point. With the estimated angle, the received image can be anti-rotated to its regular position.

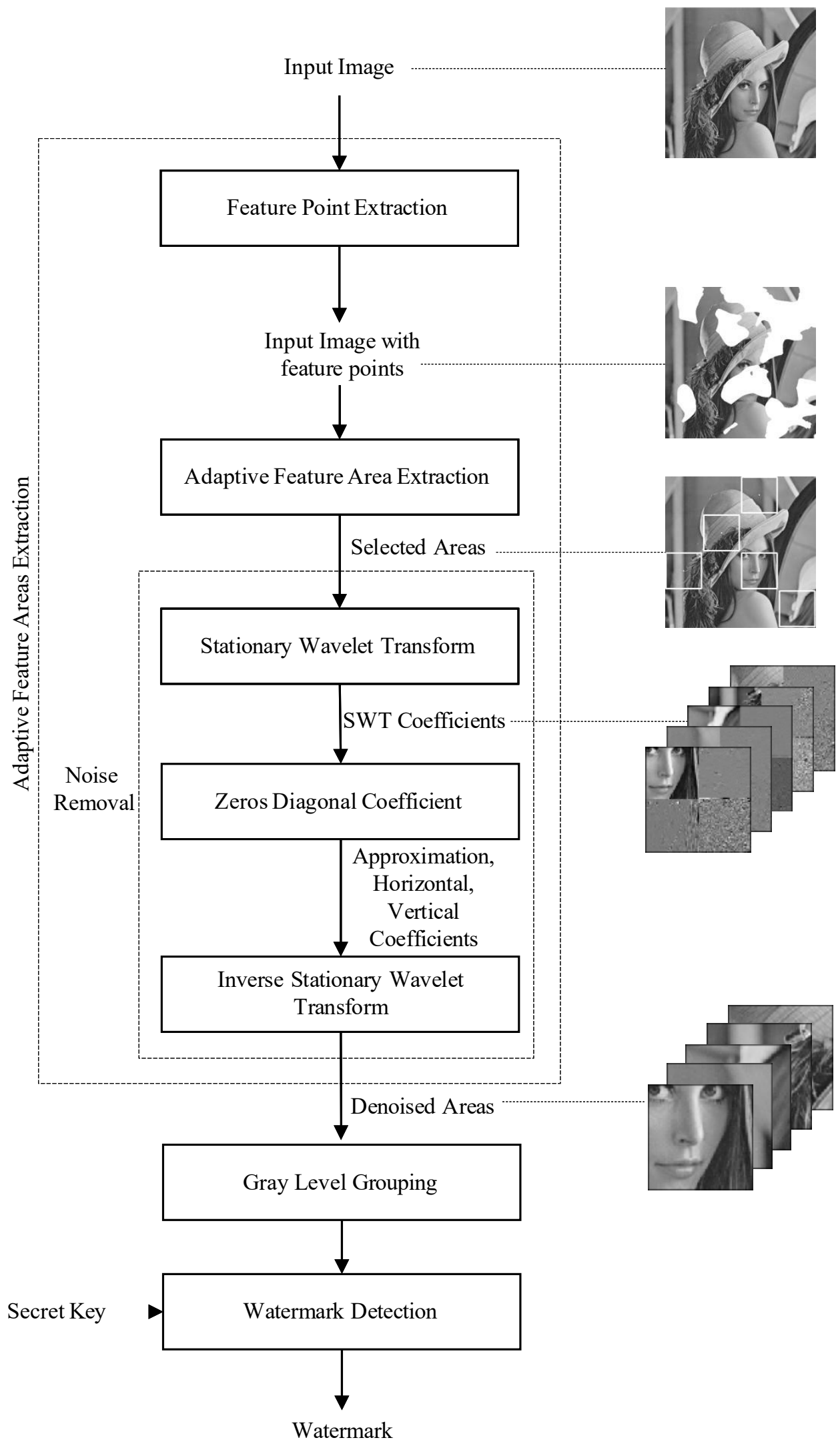

**Fig. 7** Flowchart of watermark extraction process

After the angle adjustment, we implement the watermark decoding. The watermark decoding process is illustrated in Fig.7. In this process, the whole procedure is similar to watermark embedding process. Firstly, local embedding areas are extracted from the received image. Then, SWT is applied on each area and denoised areas are obtained by iSWT procedure. After that, the gray level of histogram of those denoised areas are grouped under the same gray level grouping rule as described in section 2.2.1. Finally, the secret key is used to extract the watermark bits in the formed gray level groups of denoised areas.

We compute the number of pixels in each bin of the selected groups, extracting watermark bits by (23). Then, the watermark can be decoded.

$$\begin{cases} w' = 1, \text{ if } \frac{N_{bin}(2d-1)}{N_{bin}(2d)} \geq 1 \\ w' = 0, \text{ if } \frac{N_{bin}(2d-1)}{N_{bin}(2d)} \leq 1 \end{cases} \tag{23}$$

where $w$ indicates the decoded binary bit and it equals 1 or 0. $N_{bin}(2d-1)$ indicates the pixel number in the Bin 1 of group $d$ while $N_{bin}(2d)$ indicates the pixel number in the Bin 2 of group $d$.

# 4 Experimental results

In this section, we evaluate the perceptual quality and the robustness of the watermarked image under different attacks in comparison with several state-of-art watermarking methods. The capacity of our method is also been compared with other watermarking methods. Here we test many digital images from the standard grayscale test images databases (http://decsai.ugr.es/cvg/CG/base.htm and http://decsai.ugr.es/cvg/dbimagenes/g256.php). And some of them are shown in Fig. 8. With PC of 64-bit operating system and Intel(R) Xeon(R) E5-2630 v3 @ 2.40GHz CPU, the execution time for watermark embedding and extraction are 45.071850 seconds and 40.181407 seconds respectively in average. The experiment results of proposed method are illustrated in Table 1 and Table 2, comparing with the state-of-the-art watermarking methods.

The maximal capacity $C = L_w * S_n$, where $L_w$ indicates the maximal length of watermark in each local area and $S_n$ means the maximal number of local regions. In the proposed scheme, to reduce the modification of the histogram during the histogram shifting and make the distribution of the modified histogram smoother, six gray levels are gathered into a bin, and two bins are formed into a group to embed one watermark data bit, as illustrated in Fig. 5. In total, 12 gray levels are chosen to embed one bit, thus for grayscale images, maximally $\lfloor 256/12 \rfloor = 21$ bits can be embedded, that means, in each local area, the maximal watermark capacity is 21. With the proposed feature area extraction method, the maximal number of local regions is 16. Therefore, $L_w = 21$, $S_n = 16$ thus $C = 336$. In our experiments, we define the number of embedded watermark bits in each local area as $N_L < \lfloor L_w/2 \rfloor$, therefore, $N_L = 9$. To make the proposed method comparable with the existing methods, the total number of watermark bits should be between 35˜50 bits, in this way, five local areas are selected for watermarking and the length of watermark is $N_{W\ at} = 45$. Fig. 9 illustrates the watermarked image under different attacks. In our experiments, we test the properties of the watermarked mage under different attacks. Because of the local embedding areas, the watermarked image cannot resist to high percentage cropping attack which could cause local embedding areas completely lost. Thus 10% and 20% cropping attacks are tested in our experiment which would cause strips lost in the watermarked image.

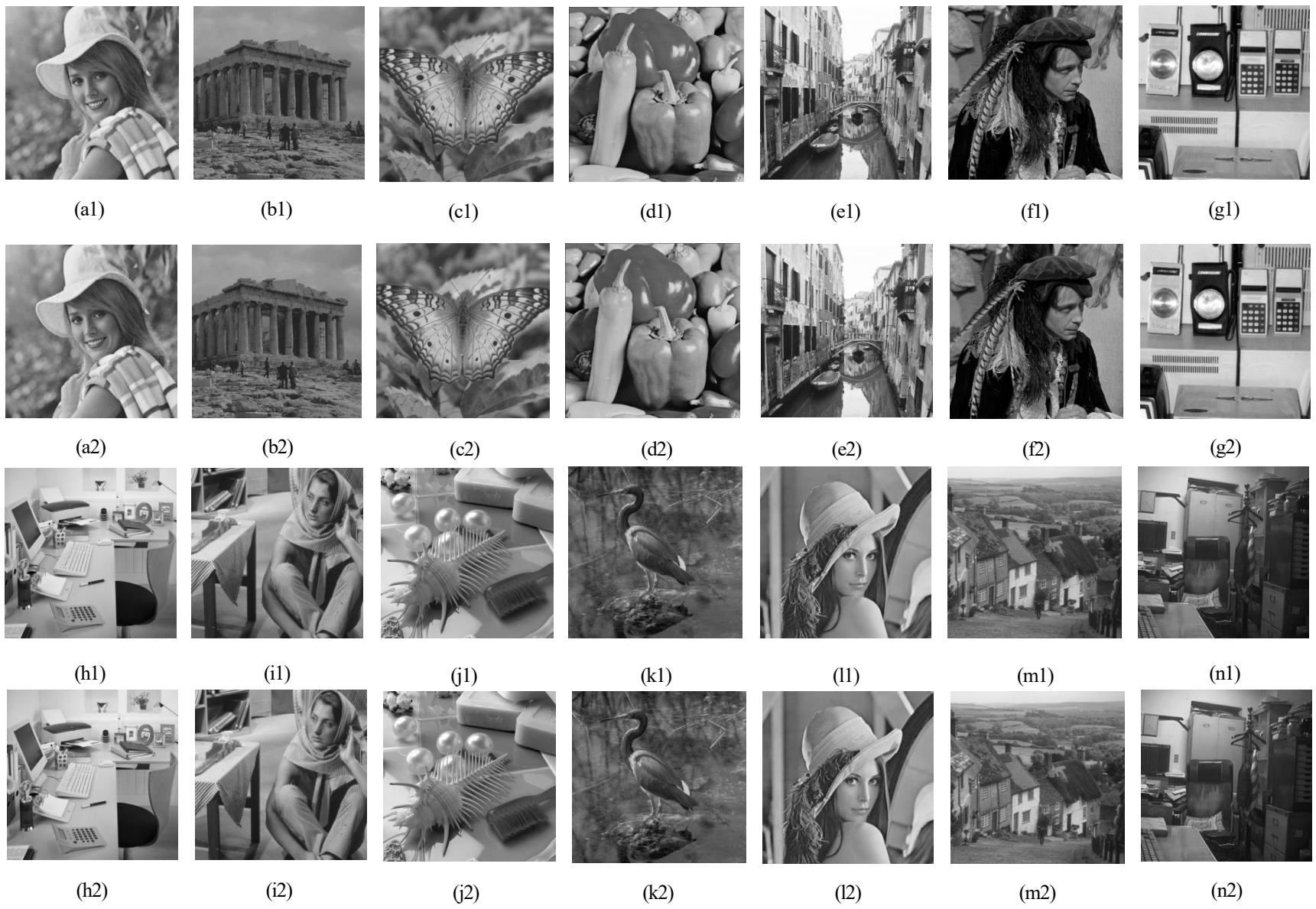

**Fig. 8** Test images from standard grayscale images databases. (a1-n1) are original images, (a2-n2) are watermarked images of (a1-n1)

The watermark can be detected correctly without any attacks on the watermarked image. The BER is counted when the watermarked image is processed under different attacks. We can see from Fig. 9 that the BER is very low under common image processing and noise attacks. The BER can even reach to zero when the watermarked image goes through cropping and median filtering.

Fig. 10 exhibts the BER line charts of the proposed watermarking method under different attack parameters and watermark strength. All the line charts show the graduate trend of performance under a specific range of attacks. Although the BER becomes higher when the attack strength becomes more intensive, the whole performance under these attacks is still good. BER is also related to watermark strength. As we can see in the line charts, the BER is slightly reduced when the watermark strength becomes higher especially

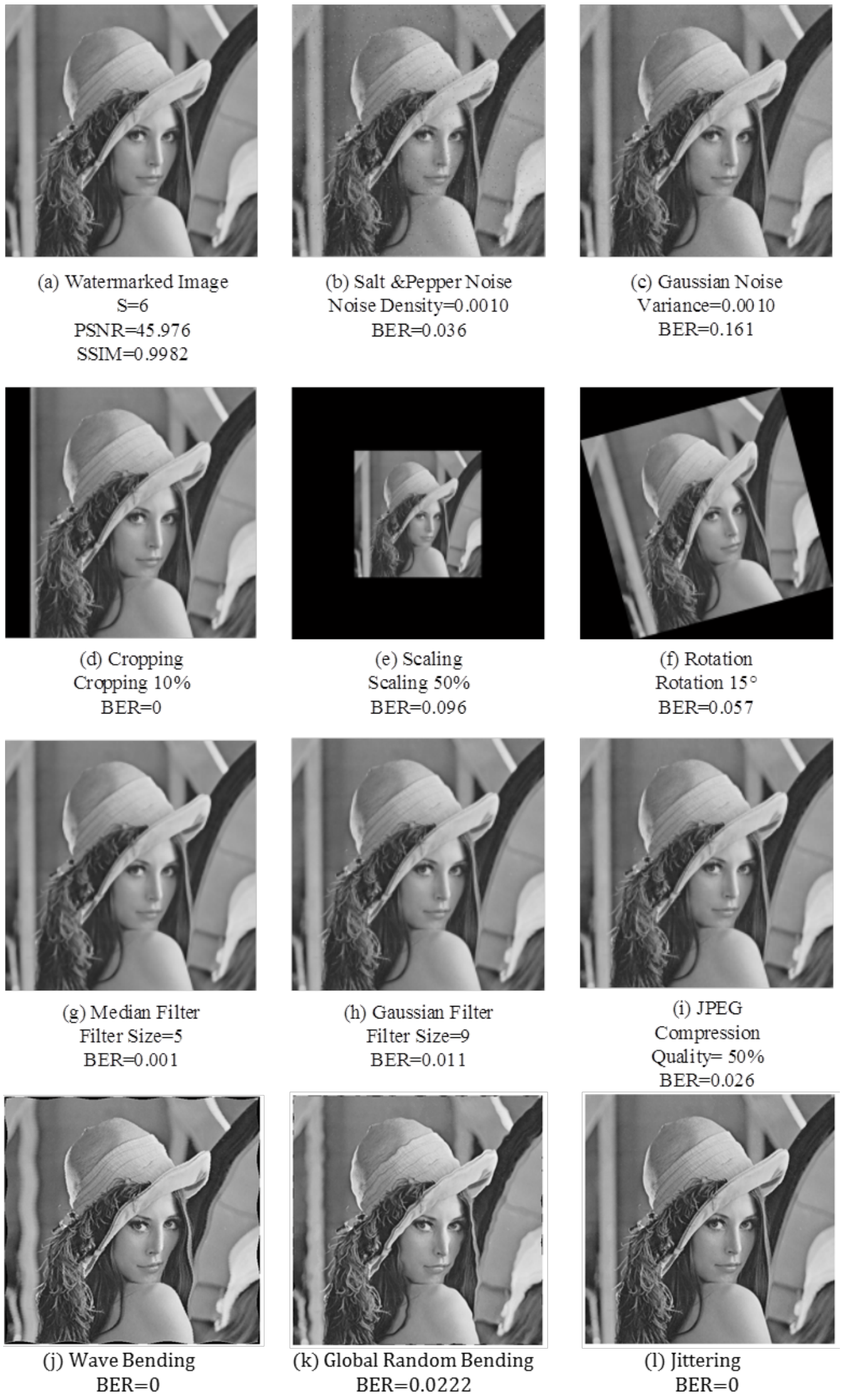

**Fig. 9** Illustration of attacks: (a) Watermarked image; (b) Salt & Pepper Noise addition; (c) Gaussian Noise addition; (d) Cropping; (e) Scaling; (f) Rotation; (g) Median Filtering; (h) Gaussian Filtering; and (i) JPEG Compression; (j) Wave Bending; (k) Global Random Bending; (l) Jittering

under the attacks of rotation and JPEG compression. But this phenomenon is less obvious under the attack of Gaussian noise and scaling. And the line of each watermark strength shows almost the same BER in the range of attack strength.

The proposed method shows good performance of resisting to common image attacks such as median filter,Gaussian filter, salt and pepper noise and rotation in which the BER is not greater than 0.09 even under the high attack intensity. The performance of proposed method also shows good robustness under JPEG compression and the BER is not greater than 0.2 even the compression quality reaches to 20%. Fig. 11 indicates the watermarked image quality under different watermark strength. Two line charts show a decreasing trend when the watermark strength becomes greater. However, the PSNR and SSIM are still in a good range and not less than 45.5 and 0.998 respectively which illustrates good performance and decent image quality of our proposed method.

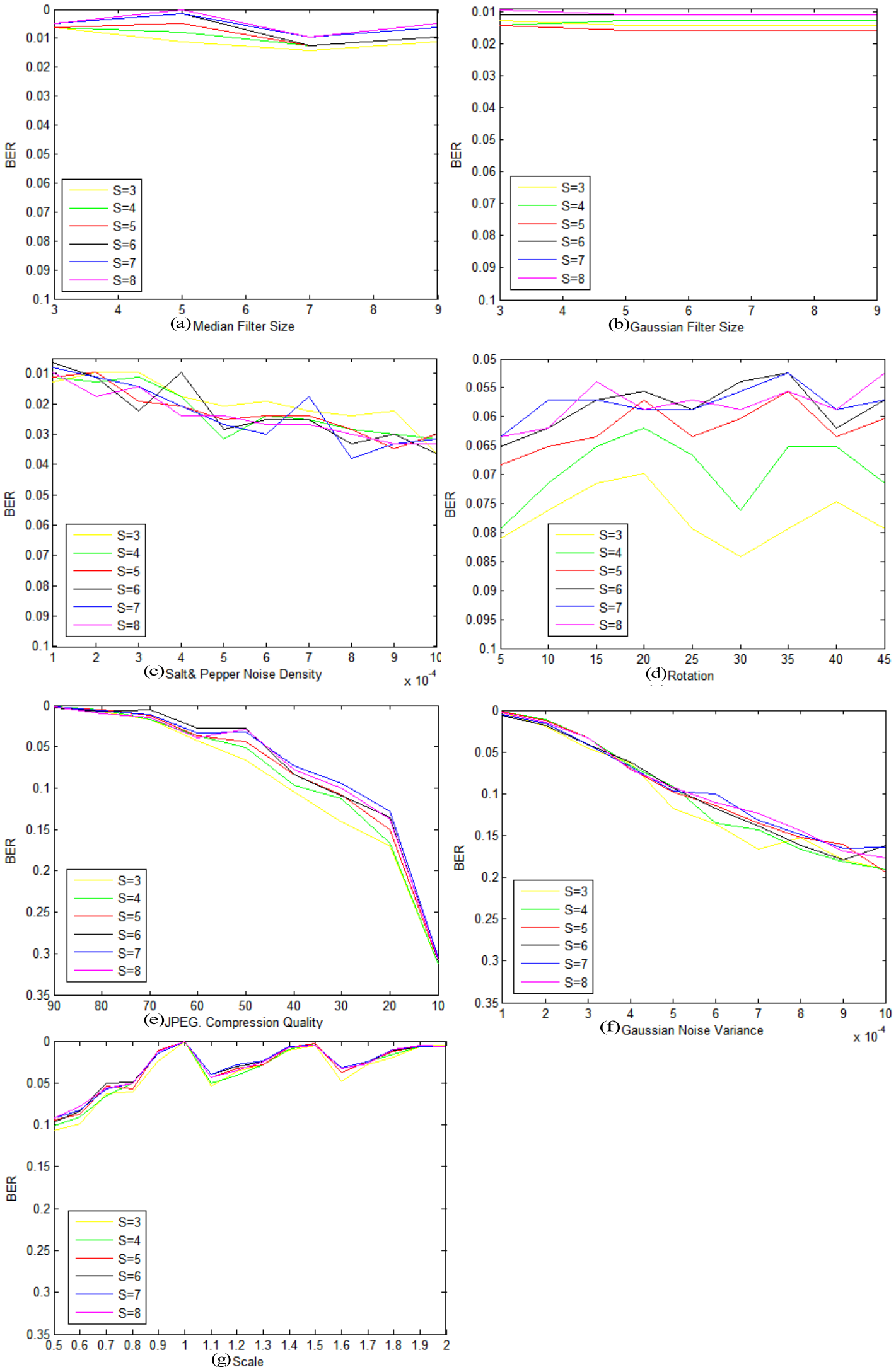

**Fig. 10** BERs of proposed method under different attacks: (a) Median Filter; (b) Gaussian Filter; (c) Salt&Pepper Noise; (d) Rotation; (e) jpeg Compression; (f) Gaussian Noise; (g) Scale.

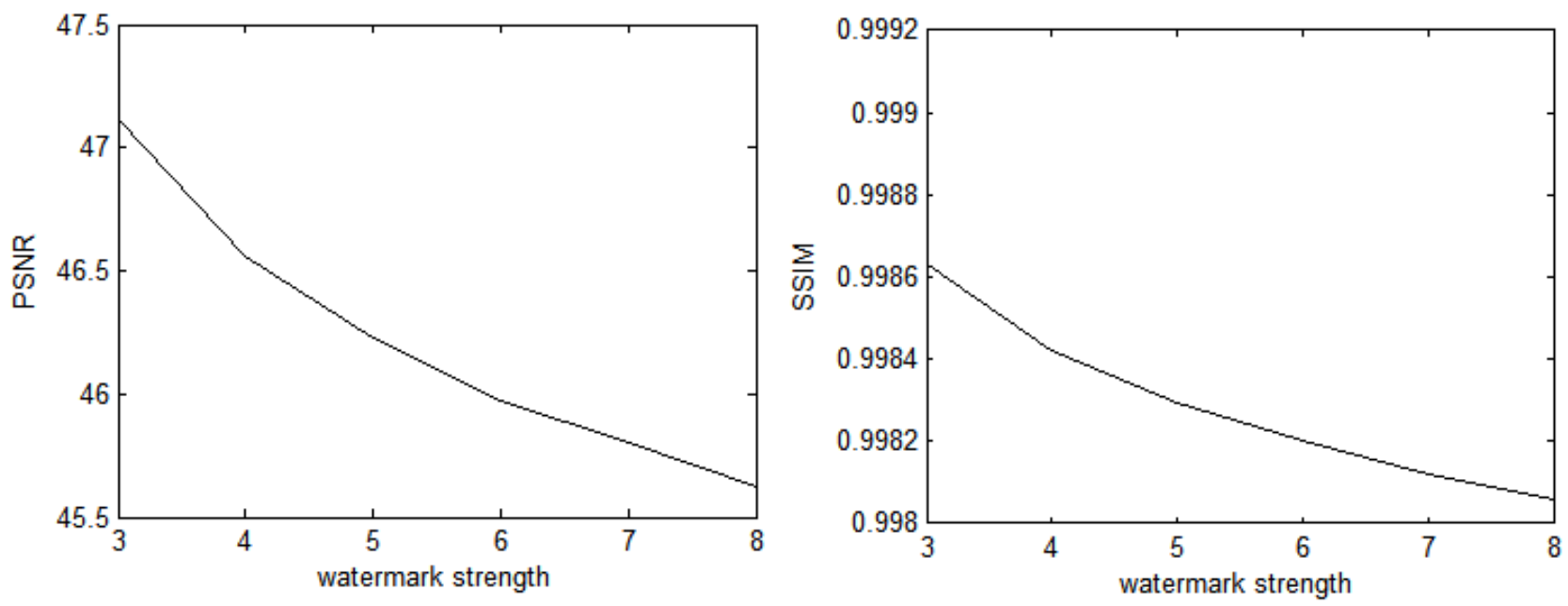

**Fig. 11** PSNR and SSIM value under different watermark strength

Table. 1 shows the experiment results of proposed method under the watermark strength of six and BER comparison with other watermarking methods.

**Table 1** BERs of the proposed method and the method in [28, 30–32]

| Attack BER | Xiang [31] | Li [28] | Deng [32] | Zong [30] | Proposed method |
|---|---|---|---|---|---|
| Gaussian noise | 0.096 | 0.476 | 0.189 | 0.011 | 0 |
| Median filtering | 0.112 | 0 | 0.121 | 0.032 | 0.004 |
| Salt& pepper noise | 0.066 | 0 | 0.086 | 0.004 | 0.022 |
| Gaussian filtering | N | N | 0.516 | N | 0.011 |
| Cropping 10% | 0.080 | 0 | 0.073 | 0.004 | 0 |
| Cropping 20% | 0.160 | 0 | 0.083 | 0.005 | 0 |
| Scaling 80% | 0.093 | 0.297 | 0.249 | 0.023 | 0.052 |
| Scaling 120% | 0.080 | 0.241 | 0.226 | 0.013 | 0.033 |
| Scaling 150% | 0.010 | 0.137 | 0.322 | N | 0.004 |
| Rotation 2° | 0.140 | N | N | N | 0.060 |
| Rotation 5° | 0.180 | 0.150 | 0.483 | N | 0.063 |
| Rotation 10° | 0.091 | 0.268 | 0.237 | 0.021 | 0.060 |
| Rotation 25° | 0.106 | 0.310 | 0.251 | 0.025 | 0.055 |
| Rotation 30° | 0.090 | 0.100 | 0.548 | N | 0.052 |
| Rotation 45° | N | 0.112 | N | N | 0.055 |
| JPEG 90% | 0.050 | N | 0.064 | N | 0.003 |
| JPEG 70% | N | N | 0.080 | N | 0.004 |
| JPEG 50% | N | N | 0.129 | N | 0.026 |
| JPEG 30% | N | N | 0.147 | 0.012 | 0.103 |
| Wave Bending | N | N | N | 0.057 | 0.026 |
| Global Random Bending | 0.054 | N | N | 0.035 | 0.046 |
| Jittering | 0.021 | N | N | 0.039 | 0.009 |

*N means there is no test in such experiment.

As we can see from the table, our proposed watermarking method is better than the method proposed by Li [28], Xiang [31] and Deng [32] generally. The proposed method shows good robustness against image attacks, reaching low BERs especially under the attacks of Gaussian noise, median filtering, cropping as well as JPEG compression. Compared with the global watermarking method proposed by Zong [30], our local embedding method shows advantages against the attacks such as Gaussian noise and median filter. According to the results, the proposed method can achieve lower BER under global random bending and jittering comparing with Xiang [31] and performs better under wave bending and jittering comparing with Zong [30]. The comparison results indicate the good performance in terms of robustness under the RBA attacks and jittering attack. Because of the proposed histogram shifting method, more gray levels in the histogram are used for watermark embedding so that the robustness of the modified histogram can be achieved and watermark can be decoded in a lower BER after bending and jittering attacks. Since the local embedding areas are included in the whole image, the BER of watermarked image under cropping attacks is zero if the cropped area does not affect the local areas. Even the cropping area affects the local embedding areas but not crop the local embedding area completely; our method can still hold a good performance under such attack. According to our experiments, we have tested the cropping of percentage up to 20% (starting from the upper-left corner, 20% of the test images are cropped), and the watermark can be successfully extracted in the test scenarios. This happens because when we extract the feature areas, the feature areas nearby corners and edges are filtered. In terms of large percentage cropping when the feature areas will fail to be extracted, the watermark extraction will fail accordingly. Because of our local method, the embedding areas are smaller than global area which would be the disadvantage to conquer some attacks (e.g. scaling and rotation) and keep the information correctly compared to global embedding method. We can see the BER in Table.1 that the BERs of the proposed method are generally low and in the good range of detectability. Thus, our method has a good robustness against a variety of attacks.

**Table 2** Property of the proposed method and the method in [28, 30–32]

| method property | Xiang [31] | Li [28] | Deng [32] | Zong [30] | Proposed method |
|---|---|---|---|---|---|
| Capacity | 25 bits | 40 bits | 20 bits | 30 bits | 45 bits |
| PSNR | 48 dB | 46.5 dB | 55 dB | 47 dB | 46 dB |

Table.2 exhibits the capacity and image quality of proposed method and other watermarking methods. We can see the general trend is that more binary bits embedded into the image, less PSNR we can get. Compared with the watermarking method proposed in [28], [30], [31] and [32], our method have a larger capacity due to our local embedding areas selection. The more local areas selected for watermark embedding, the larger capacity of the whole image since each local area has the ability of containing one bunch of binary bits of watermark. In terms of the image quality, our method can reach the PSNR not less than 45.5 according to the watermark strength and capacity. This means good perceptual quality can be reached while maintaining the large capacity as well as the robustness.

# 5 Conclusions

The local watermark embedding method has been implemented in many situations for digital image authentication. However, embedding watermark directly in spatial domain is relatively fragile against various image processing attacks. By using the feature extraction and denoising method, the proposed method in this paper embeds watermark into the smooth part of the local areas of the image which ensures the security and robustness of the watermark. This method enhances the imperceptibility, security as well as robustness due to the histogram shifting method, secret key and the SWT denoising process. From the experiment results we can see that, the proposed method has better robustness against attacks such as Gaussian noise, median filtering and jittering, etc. when compared with other watermarking methods. The BER line charts also indicate the watermarking performance of proposed method under different attacks and shows decent robustness in such range of attack intensity. Since our method is implemented in local areas of the whole image and due to the reconstruction procedure, the PSNR of the watermarked image can be improved even under a high watermark embedding strength. Thus, the proposed watermarking method can not only reach a good performance in case of robustness but also has a good image quality of the watermarked image.